\documentclass[sigconf]{acmart}

\copyrightyear{2023}
\acmYear{2023}
\setcopyright{rightsretained}
\acmConference[ITiCSE 2023]{Proceedings of the 2023 Conference on Innovation and Technology in Computer Science Education V. 1}{July 8--12, 2023}{Turku, Finland}
\acmBooktitle{Proceedings of the 2023 Conference on Innovation and Technology in Computer Science Education V. 1 (ITiCSE 2023), July 8--12, 2023, Turku, Finland}\acmDOI{10.1145/3587102.3588787}
\acmISBN{979-8-4007-0138-2/23/07}

\makeatletter
\gdef\@copyrightpermission{
  \begin{minipage}{0.3\columnwidth}
   \href{https://creativecommons.org/licenses/by/4.0/}{\includegraphics[width=0.90\textwidth]{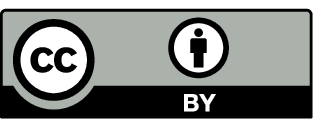}}
  \end{minipage}\hfill
  \begin{minipage}{0.7\columnwidth}
   \href{https://creativecommons.org/licenses/by/4.0/}{This work is licensed under a Creative Commons Attribution International 4.0 License.}
  \end{minipage}
  \vspace{5pt}
}
\makeatother

\begin{document}

\title{Automated Questions About Learners' Own Code Help to Detect Fragile Knowledge}

\author{Teemu Lehtinen}
\email{teemu.t.lehtinen@aalto.fi}
\orcid{0000-0003-4794-3818}
\affiliation{%
  \institution{Aalto University}
  \city{Espoo}
  \country{Finland}
}

\author{Otto Seppälä}
\email{otto.seppala@aalto.fi}
\orcid{0000-0003-4694-9580}
\affiliation{%
  \institution{Aalto University}
  \city{Espoo}
  \country{Finland}
}

\author{Ari Korhonen}
\email{ari.korhonen@aalto.fi}
\orcid{0000-0002-2784-7979}
\affiliation{%
  \institution{Aalto University}
  \city{Espoo}
  \country{Finland}
}

\renewcommand{\shortauthors}{Teemu Lehtinen, Otto Seppälä, \& Ari Korhonen}

\begin{abstract}
Students are able to produce correctly functioning program code even though they have a fragile understanding of how it actually works. Questions derived automatically from individual exercise submissions (QLC) can probe if and how well the students understand the structure and logic of the code they just created. Prior research studied this approach in the context of the first programming course. We replicate the study on a follow-up programming course for engineering students which contains a recap of general concepts in CS1. The task was the classic rainfall problem which was solved by 90\% of the students. The QLCs generated from each passing submission were kept intentionally simple, yet 27\% of the students failed in at least one of them. Students who struggled with questions about their own program logic had a lower median for overall course points than students who answered correctly.
\end{abstract}




\maketitle

\section{Introduction}

While novice programmers might be able to produce a working solution to a programming problem, some of them might struggle and answer incorrectly even simple questions about their own (functionally correct) program code~\cite{lehtinen2021students}. Previous research suggests that this phenomena may be due to fragile programming skills in general~\cite{perkins1986fragile}, misconceptions on programming constructs~\cite{madison2002modular, simon2011assignment}, inability to trace code~\cite{lister2004multinational, cunningham2020not} or that the failing students did not construct the program by themselves~\cite{vogts2009plagiarising}. 
Recent advances in AI are likely to exacerbate the problem by supplying code that the student may not fully comprehend. All of the previous reasons can seriously hinder learning on a second programming course.

We partially replicate previous studies on automatically generated \emph{Questions about Learners' Code} (QLCs)~\cite{lehtinen2021students, santos2022jask, lehtinen2023automated} by adapting selected research questions to new context. The previous studies focused on students' program comprehension during the first weeks of their \emph{first programming course} (CS1). The studies found that a number of students perform poorly when answering questions on their own functionally correct code. In addition, two studies reported that students who answer incorrectly to QLCs have on average lower success on the course.

The three papers call for replications and improved research designs as future efforts. In the present study we answer this call: research QLCs in a new context, a programming course for non-CS majors with a CS1 as prerequisite. As the students are further in their programming studies we expect them to have better programming skills and ability to answer QLCs similar to those in the previous studies. We do not, however, have access to their grades or research data about their performance on the prerequisite course.



We utilize QLCs on a \emph{second programming course} to collect cases where students submit a program, but do not properly understand how it works. A variation of the classic Rainfall-problem~\cite{soloway1984novices} was used as a programming recap exercise. The students submitted their solutions to an automated grading platform allowing resubmits. 
After the programming task, students were offered multiple-choice QLCs that were generated for the program they submitted and which we expected the programmer should be able to answer. The students' answers to these questions were analyzed to discuss the following research questions:

\begin{description}
\item[RQ1] How well do students answer QLCs for an exercise targeting prerequisite knowledge?
\item[RQ2] How do the QLC results relate to success on second programming course?
\end{description}

We hypothesized that QLCs could reveal weaknesses in prerequisites which again can lead to low grades or even drop-outs. We found that success rates for the different types of QLCs were comparable to those reported in previous studies~\cite{lehtinen2021students, santos2022jask, lehtinen2023automated} even though the students had passed a CS1 in Python. Furthermore, 
students answering incorrectly about their own program logic had a lower median for course points compared with those that answered correctly.




The rest of the paper is structured as follows. 
Section~\ref{sec:related} gives some background and introduces the related studies of which selected questions are replicated in this research. 
Section~\ref{sec:method} gives the context in which this study is conducted, introduces the exercise used in the study, and explains how the data was analysed. 
Section~\ref{sec:results} summarises the results, and finally 
Section~\ref{sec:discussion} discusses the findings, their interpretations, and some threats to validity. 

\section{Related Work}
\label{sec:related}

\emph{Program comprehension} (PC) has been researched extensively. \linebreak \citet{schulte2010introduction} review PC literature from a computer science education research perspective. They map existing PC models to a proposed \emph{revised block model} for education. The block model describes PC in two dimensions: a scale of atoms--blocks--relations--macro and other dimension on text--execution--purpose~\cite{schulte2008block}. Those dimensions combine to areas or a diagram of blocks where we can examine comprehension; for example, \emph{execution} through a sequence of \emph{related} function calls.

Although many people often think of programming with focus on writing code, researchers have recognized the importance of teaching program comprehension to novices and many learning activities are suggested and developed to address this issue~\cite{izu2019fostering, izu2020comparing, mirolo2020highschool, shargabi2020performing, kumar2021refute}.
Another common conception is that if a person writes a program they can reason how it works. However, this is not always true and cases where students have poor comprehension of the programs they wrote have been reported~\cite{madison2002modular, kennedy2019coding, lehtinen2021students}.

\citet{perkins1986fragile} researched difficulties that novice programmers encounter. In their experiment, instructors worked one-on-one with the programming students and asked them questions such as what does this element do? While such prompts help students go forward the students' reactions also reveal cases of fragile knowledge. They further report on different types of fragility they identified.

\citet{lehtinen2021lets} defined QLCs as an approach that poses questions to a student about the concrete structures and patterns in a program, which the same student previously created. They hypothesize such questions could test different areas of PC, catch cases of unproductive success, and trigger reflection including self-explanation. Furthermore, they describe a potential design of a system to automate generation of QLCs.

\citet{lehtinen2021students} report on an early pilot for QLCs where students wrote open-text answers to QLCs that were manually prepared for selected program writing exercises for novices. Their main research questions are ``How well do students answer QLCs?'' and ``How success in QLCs correlates with other learning data?''. They used the block model to design their QLCs and report the following student success rates: atom--text 95\%, relation--text 67\%, atom--execution 72\%, block--execution (80\% and 75\%), macro--execution (58\% and 79\%), and macro--purpose 91\%. Furthermore, among the students who created a functionally correct program the ones who answered the related set of the first three QLCs correctly had increased course success and retention in comparison to students who answered incorrectly.

\citet{santos2022jask} present a tool that automates QLCs for Java programs and report on a pilot evaluation on CS1. The evaluation asks ``How do students perform in QLCs in their lab exercises?'' and ``How do students perceive the activity of answering QLCs?''. They measure above 80\% success rates for QLCs targeting static aspects (text in block model) and below 50\% success rates for QLCs targeting dynamic aspects (execution in the block model). Despite the low success in the latter, students reported good average confidence levels in both static and dynamic cases. The authors also note that success rates have large variations for identical QLC templates when the programming task changes, in other words, difficulty is dependent on the program. In a post-exercise survey majority of the students reported learning or reinforcement of their knowledge regarding terminology. Approximately one third of students reported the same regarding loops, recursion, and variable values.

\citet{lehtinen2023automated} present automated QLCs for JavaScript programs as well as pilot study on students' answers. They report success rates all the way from 33\% up to 97\%. In the upper end the QLCs target block model's text level and in the lower end they require tracing in block model's execution level. They find that students who repeatedly answer QLCs incorrectly also have more challenges while writing the program and tinker towards solutions. Respectively, the average error rate in QLCs correlates negatively with course success.

\section{Method}
\label{sec:method}


\subsection{Context}

This study was conducted in a 12 weeks 5 ECTS\footnote{The European Credit Transfer and Accumulation System} programming course given in Aalto University, a large research university in Finland. The course includes topics such as introduction to Numpy and Scipy modules in Python, basics in databases and how to interconnect Python and SQL, MATLAB and Simulink programming, and simple web programming with Javascript.
The course has one 5 ECTS prerequisite programming course in Python. QLCs were designed to be part of a recap exercise in week 1, which also included three other programming exercises and a couple of multiple choice questions. All of the exercises were automatically assessed and the students got immediate feedback on their solutions. In case the student did not achieve full points on the first attempt, 9 resubmissions were allowed.  

The course had no final exam, but the final grade was calculated based on the points received from the exercises and a project done in groups (max. 3 students). To pass the course, students needed to do at least half of the individual weekly exercises and the project. To support this, the course had lab sessions on every weekday for seeking help from teaching assistants (TA). The course had also 5 live lectures and a couple recorded video lectures. Due to the COVID-19 situation, the live lectures and lab sessions were in Zoom\footnote{Zoom is a communications platform that allows users to connect with video, audio, phone, and chat over Internet.}. 

Typically some 350 engineering students\footnote{Students majoring e.g. from Mechanical and Civil Engineering, but not Computer Science} start doing the course each year. There are no penalties for dropping the course and only about 75-80\% pass the course yearly. Most of the students are first year students, but many might end up taking the course on the second or third year, which also explains the relatively high drop out rate. 

The study was conducted in Spring 2021. 324 students volunteered to participate and gave their consent on the online course platform. While all students did the same tasks, the volunteering students allowed their anonymous exercise data (timestamps, answers, automated feedback, points) to be used in our research. We offered no incentives for the participation and informed that participation had no effect on the course requirements or assessment. The course instructor was one of the second authors. The local regulations did not require an ethical board's review for our research.

\subsection{Task}

The first week of the course included Python recap-material. We selected the first program writing exercise, called ``Calculation of average rainfall'', for our QLCs experiment. For the rest of the course and the 98 other exercises we only calculate students' exercise points. The rainfall task was to create a program according to the following description which we translated to English for this article:

\begin{quotation}\small\emph{
``Implement function rain(), which asks the user for days' rainfalls and calculates the average rainfall for the given days. The program ends by providing -999 as input.  Notice that rainfall cannot be negative so your program must ignore negative numbers. The program must also ignore other inputs beside numbers (e.g. letters). If the user does not input any acceptable numbers the program must return 0.''}
\end{quotation}

In addition to the description, an example of expected input and output lines was given. The provided code template included an empty function stub and a ``main'' block.

The students submitted their program online and received automated feedback based on four unit test cases:
\begin{enumerate}\small
\item the program ends when -999 is given as input
\item the program does not fail when lines of letters is given as input,
\item the program ignores input that cannot be transformed into integer numbers,
\item the program outputs the average of the numbers given as input.
\end{enumerate}
The differences from the expected output were displayed to the student. In addition, TAs were available to answer questions and give hints during online lab sessions. Based on the number of passed unit test cases students received up to 95 points form the program writing task. Students could resubmit their program up to 9 times to receive new assessment and points.

Once the students had submitted their program at least once they could open a questionnaire that included 2--3 generated QLCs. If the program did not include a structure necessary to create one of the question types the student would not receive that question type. Students could only answer the QLCs once and received 5 points if they answered correctly. Together the program writing and QLCs were worth 100 points which were 1.4\% of the total available on the course.


QLC generation uses question templates and static program analysis as described in \cite{lehtinen2021lets}. The library for generating QLCs for any python code is available as open source software\footnote{\url{https://github.com/teemulehtinen/qlcpy}} and contains both documentation and code examples.
The selection of QLC types has been further extended from the types included in this study. 

\begin{figure}
    \centering
    \includegraphics[width=7.45cm]{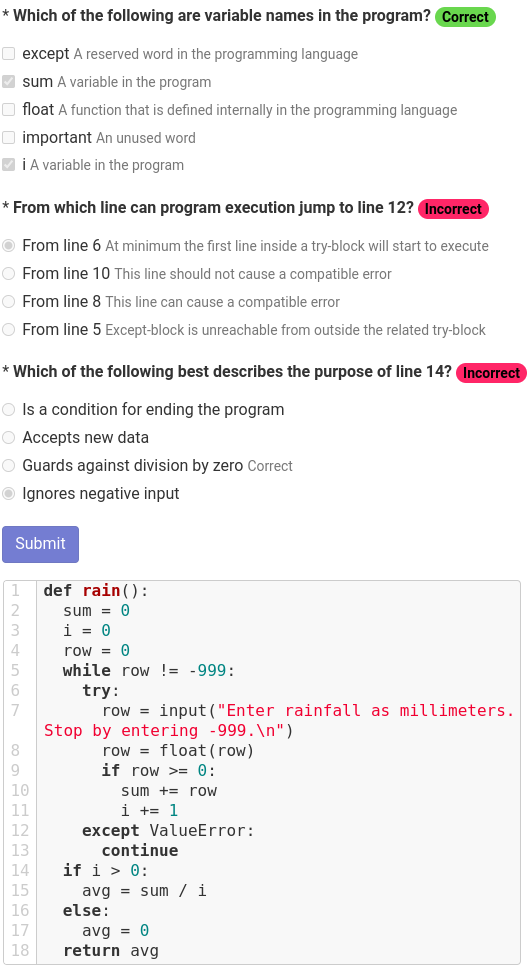}
    \caption{An example of a generated and answered questionnaire for a program that completes the rainfall task.}
    \label{fig:ui}
\end{figure}

\autoref{fig:ui} presents an example questionnaire that was generated for the included program and then answered by the user. These were multiple-choice questions and students could only answer once. For increased learning, descriptions for each option were displayed after the user had submitted their answers.

\autoref{tab:answers} presents the selected QLC templates (Q1-Q3) and descriptions of their correct answer options as well as their incorrect distractor options. The QLCs were designed to approach different areas of program comprehension. We situate these questions in the two dimensions of the block model for program comprehension.

Q1 has been a typical question type in previous QLC research. It tests the students' recognition of an atomic element, in this case a variable, in the program's structure (atom--text). 

Q2 asks from which lines can the execution enter an except-block. While testing understanding of block-structures, it relates the expected error to the potential source of the error (relations--execution).

Q3 takes a step deeper into the purpose of the code. More specifically, it asks for the purpose of a statement in its context (atom--purpose). The answer options are designed so that students should have some understanding of the program's design to answer correctly.



\subsection{Analysis}
\label{sec:analysis}

Anonymized data was extracted from the course platform for those 324 students that gave a research consent. We rejected 33 students who did not create the rainfall program or did not answer the related QLCs. Finally, we accepted 291 students (90\%) in our research and analysed their exercise data quantitatively.

To answer RQ1, we counted the numbers of correct as well as incorrect answer options. For each type of question, we calculated success rates as the number of correct answers divided by the total number of answers. We discuss the relative difficulty of the different QLC types in this study and compare to success rates reported in earlier research.

To measure success on course we use \emph{Course Points} that is the sum of exercise points student earned during the course. The course grade was assigned based on these points as well.  To answer RQ2, we separately investigate the three types of QLCs we generated. For each QLC type, we compare the students who answered correctly to those who answered incorrectly.

According to visual inspection, the course points have non-normal distribution.
The maximum points available on the course cause a ceiling effect as the automatically assessed exercises can be submitted 10 times to get feedback and collect more points. Also course points tend to accumulate close to the limits required for different course grades.

In case of non-normal distributions, we calculate Mann-Whitney $U$-tests for the hypothesis that two groups, students answering QLCs correctly and incorrectly, have equal medians. In addition, we calculate common language effect size. We use an alpha level of $p < .05$ to reject the hypothesis and as we test three comparisons we use Bonferroni correction $p / 3 = .017$ to avoid false-positive analysis.

\section{Results}
\label{sec:results}

\begin{table}[t]
\centering
\caption{The three QLCs, descriptions of their correct and incorrect answer options, and numbers of students who answered each option. \normalfont{Success rates are in bold text and indented lines list student counts for actual labels used as answer options.}}
\begin{tabular}{lcr}
\toprule
\multicolumn{3}{l}{{\bf Q1 Which of the following are variable}} \\
{\bf names in the program?} ($N=291$) & \multicolumn{2}{c}{Students} \\
\midrule
\textbullet~ {\bf Correct options selected} & {\bf 249} & {\bf 86\%} \\
\textbullet~ Missed a variable & 22 & 8\% \\
\hspace{0.5cm} i (6)\, list (4)\, sum (2)\, {\it one-off names} (11) &&\\
\textbullet~ Selected a built-in function & 21 & 7\% \\
\hspace{0.5cm} print (7)\, float (7)\, input (6)\, len (1) &&\\
\textbullet~ Selected a reserved word & 20 & 7\% \\
\hspace{0.5cm} return (6)\, if (5)\, while (3)\, try (2) &&\\
\hspace{0.5cm} except (2)\, pass (1)\, for (1) &&\\
\textbullet~ Selected an unused word & 3 & 1\% \\
\hspace{0.5cm} n (1)\, total (1)\, other (1) &&\\
\midrule
\multicolumn{3}{l}{{\bf Q2 From which line can program}} \\
{\bf execution jump to line X?} ($N=243$) & \multicolumn{2}{c}{Students} \\
\midrule
{\it The line in question starts an except-block.} & 243 &\\
\textbullet~ {\bf Correct line} & {\bf 207} & {\bf 85\%} \\
\hspace{0.5cm} int/float (196)\, input (7)\, division by zero (4) &&\\
\textbullet~ The related \texttt{try} line & 33 & 14\% \\
\textbullet~ Outside and before the related try-block & 3 & 1\% \\
\midrule
\multicolumn{3}{l}{{\bf Q3 Which of the following best describes}} \\
{\bf the purpose of line X?} ($N=291$) & \multicolumn{2}{c}{Students} \\
\midrule
{\it The line in question calls built-in function input.} & 234 &\\
\textbullet~ {\bf Correct: Accepts new data} & {\bf 227} & {\bf 97\%} \\
\textbullet~ Ignores negative input & 4 & 2\% \\
\textbullet~ Is a condition for ending the program & 2 & 1\% \\
\textbullet~ Guards against division by zero & 1 & 0\% \\
\midrule
{\it The line in question is a condition that skips} &&\\
{\it a block where zero would divide.} & 57 &\\
\textbullet~ {\bf Correct: Guards against division by zero} & {\bf 54} & {\bf 95\%} \\
\textbullet~ Is a condition for ending the program & 2 & 4\% \\
\textbullet~ Ignores negative input & 1 & 2\% \\
\bottomrule
\end{tabular}

\label{tab:answers}
\end{table}


\begin{figure*}
\centering\includegraphics{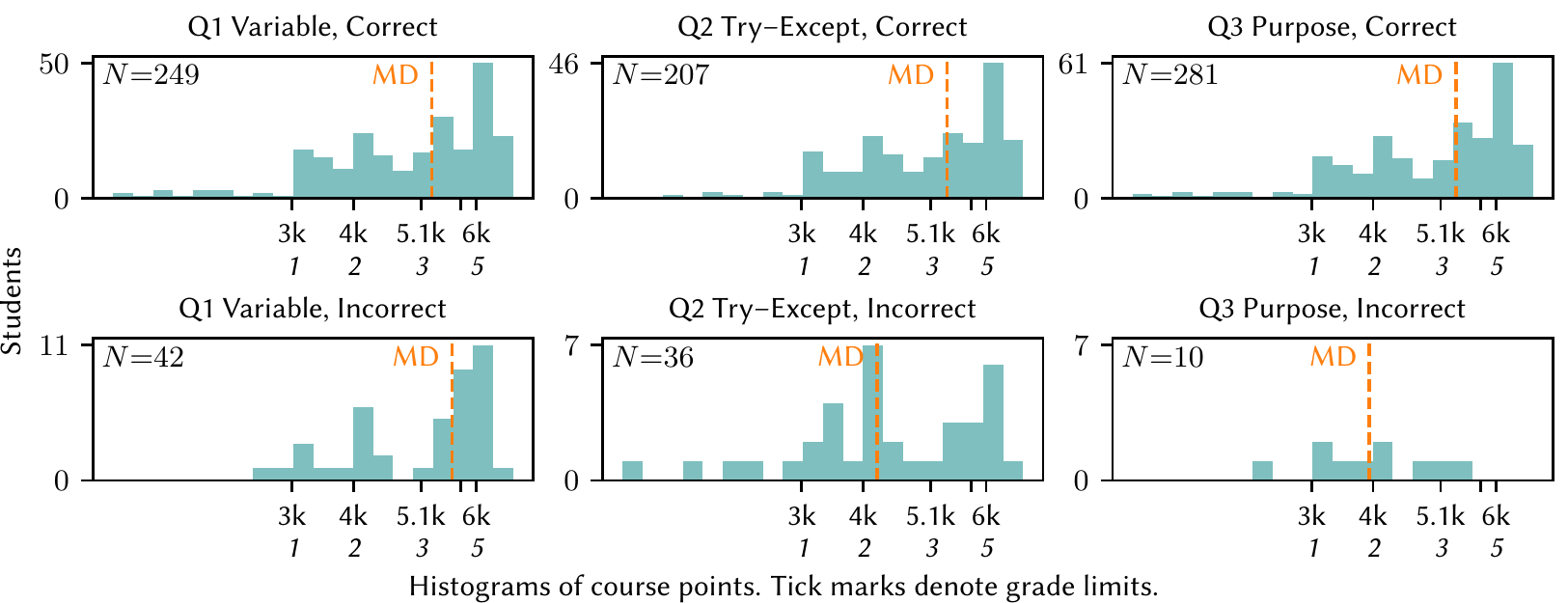}
\caption{Students' course performance in groups by whether they answered correctly or incorrectly to the QLC types.}
\label{fig:types}
\end{figure*}

\begin{table}
\centering
\caption{Median course points for students who answered different QLCs correctly (T) or incorrectly (F), numbers of students, degrees of freedom, Mann-Whitney $U$-test statistic, $p$-value for the null hypothesis that the two medians are equal, and common language effect size.}
\begin{tabular}{lcccccccc}
\toprule
QLC & $\mathop{MD}_\text{T}$ & $\mathop{MD}_\text{F}$ & $\mathop{n}_\text{T}$ & $\mathop{n}_\text{F}$ & df & $U$ & $p$ & CLES \\
\midrule
Q1 & 5272 & 5603.5 & 249 & 42 & 289 & 4932 & .556 & .47 \\
Q2 & 5366 & 4227.5 & 207 & 36 & 241 & 4716 & .011 & .63 \\
Q3 & 5348 & 3941 & 281 & 10 & 289 & 2124 & .006 & .76 \\
\bottomrule
\end{tabular}
\label{tab:tests}
\end{table}

Out of the 291 students in the study 79 students (27\%) failed at least one of the QLCs generated for their program.
\autoref{tab:answers} presents success rates and counts of incorrect answers that students have for each of the three different QLC types. Next, we highlight the findings for each type and include the question's target in the block model for program comprehension to ease comparison with previous studies.

In Q1 (atom--text), students have 86\% success rate. Indented line in \autoref{tab:answers} presents common missed variable names which is different from how those variables were actually used. Four students missed a variable name that was assigned in a \emph{for}-statement and seven students missed a variable name that they used for a reference to a list structure. We did not identify other potential patterns.

The Q2 (relations--execution) was asked from 48 students less than the other two question types as the question could not be generated for programs that did not include a try--except structure. The students who received the question have an 85\% success rate. The majority (33) of incorrect answers selected the line that declares the targeted try-block and in 24 cases of those, the correct line is the next line after the selected try-line.

In Q3 (atom--purpose), students have 96\% success rate. We did confirm that none of the incorrect answers could be defended with true arguments. Possible reasons include selecting ``Ignores negative input'' when the line has a condition to skip a block where division by zero would happen, and selecting ``Is a condition for ending the program'' when actually the line accepted new input using a prompt that mentioned how to end the program.

Among the students who submitted the program and answered the QLCs there were 13 cases where the student's final program failed at least one of the functional tests. From these students 100\% answered correctly to Q1 and 92\% to Q3. 8 of these students received Q2 and 63\% answered correctly. The overlap with students who did not fix their program to pass all tests and students who answered QLCs incorrectly is small although students with incomplete programs have relatively less generations and more incorrect answers to Q2.



In \autoref{fig:types}, we study course points distributions separately for the three different types of QLCs generated and whether the student answered correctly or incorrectly.
Our statistical tests reported in \autoref{tab:tests} reveal that students who answered Q2 or Q3 incorrectly had statistically significantly lower median course points than students who answered correctly ($p < .017$) while answers to Q1 divided students into groups having statistically identical means. We calculated medium effect size for Q2 ($.63$) and large effect size for Q3 ($.76$). The common language effect size is the probability that in a randomly selected pair the student who answered incorrectly has lower course points than the student who answered correctly.

\section{Discussion}
\label{sec:discussion}


\subsection{Interpretation and Implications}

In this study, we have introduced QLCs incorporated into an automatic assessment system that takes students' program code as input, 
generates multiple-choice questions that target the program, collects students' answers, and presents automated feedback. We have analyzed the results.
In the following, we discuss the findings in the light of the two research questions and compare them with previous studies.

\subsubsection{RQ1: How well do students answer QLCs for an exercise targeting prerequisite knowledge?}


For Q1 and Q2 about 15\% of the students had answered incorrectly. This is aligned with previous QLCs that have targeted block model's {\em text} level or simple cases of {\em execution}. For these types of QLCs students have typically had above 70\% success rates on their first programming course and in many cases 10--15\% of answers were incorrect. Lower success rates have been recorded when students have to trace, that is, mentally execute the program to answer QLCs.

Only 5\% of students failed Q3 which is comparable to the best success rates recorded in previous studies. There is only one earlier evaluation of {\em purpose} level QLC where students had 91\% success rate~\cite{lehtinen2021students}.

It is surprising that the proportional number of students answering simple questions incorrectly remains the same also after completing their first programming course. Even though the student population is different in these studies, there seems to be a trend that some students can achieve very good program comprehension already during the first couple of weeks in a first programming course, while some students still struggle after completing a whole course. The reasons must be manifold. Yet, we can use the results to explore students’ programming knowledge and program comprehension. 

In Q1, some students ignored variables that were used for lists or iteration and the difference from assigning a single value to a variable may have affected their thoughts. In Q2, several students may have intentionally selected the line just before where the error occurs which may be related to their mental model of execution. A common reason for failing one or two unit tests for the program are problems in handling unexpected input with a \emph{try}-statement. Those students received less related Q2 and struggled more to answer. Many incorrect answers to Q3 suggest that the student searched for superficial hints in the code line, such as strings of text or numerical values, without understanding the actual purpose of the line.
We argue that above are signs of fragile knowledge.

Plagiarism could increase the number of incorrect answers to QLCs as students have weaker understanding of the code they submit in comparison to students who created their own programs. 
Similarly, they may copy parts of code from examples, online sources and discussions, or occasionally directly from instructors. Given that our programming task is relatively simple and programs are small students could answer QLCs correctly also when they did not create the targeted program themself. In fact, we expect that students who mastered CS1 should be able to do so.

Currently, artificial intelligence (AI) is readily available for students to produce new code they may not understand but which solves their programming assignment\footnote{For example Github Copilot and ChatGPT are AI tools that employ successful large language models to generate program code.}. In some cases AI has been found to write better programs for typical tasks in CS1 than most students~\cite{finnieansley2022robots}. \citet{raman2022programming} argue that, due to advances in AI, changes are required to the focus, pedagogy, and assessment of program writing. They recommend greater emphasis on program testing and comprehension. We believe QLCs have potential to be part of the change.

\subsubsection{RQ2: How do the QLC results relate to success on second programming course?}

All of the researched students did submit a working rainfall-program although 13 cases still failed some functional tests. The students who then answered incorrectly to Q2 or Q3 on average had lower course success. The Q2 and Q3 were designed to target student's program comprehension at block model's execution and purpose levels respectively. Student's fragile knowledge in these areas would be a hindering factor in learning more programming content. We offer this as potential hypothesis for the lower course success.
As Q1 is a checkbox-style multiple choice question it required the students' carefully consider every answer option separately. The incorrect options listed in \autoref{tab:answers} suggest that there are probably multiple different reasons for the incorrect answers.

Students in the previous study~\cite{santos2022jask} self-reported terminology as the topic they learned the most from answering QLCs.  We assume students can reason about program execution without knowing the terminology that well. We do recognize the value of terminology for communication and continued learning. However, in our study, incorrect interpretations of execution or purpose are more related with lower course success than potential gaps in knowledge of terminology.



\subsection{Threats to Validity}
\label{sec:validity}

As students volunteered for the study there is a risk of selection bias. Of all the students who started the course, 94\% gave consent to use their exercise data without any incentives. Unfortunately, 33 students did not complete the studied task. This likely excluded lower achieving students - a possible selection bias.

It is important to note that there are many possible reasons to answer incorrectly to the QLCs that we generated. We assume that these simple questions can be answered correctly without significant extra effort if the code is already understood while producing it. However, there are reasons other than poor understanding, such as carelessness which can make the student select an incorrect option. Poor understanding in turn can be a result of fragile prerequisite knowledge or even plagiarised code with no intention to understand the code. Some students may have also sought unsupervised help to answer their personal QLCs online.


We chose to have only three simple questions to keep the study short to ensure the participants stayed motivated. The students' programs were manually examined to ensure there were no valid arguments to choose an option designed to be incorrect. However, we did not interview the students to discover why they answered as they did.

In our analysis we use non-parametric statistical tests which are less powerful than $t$-test for normal distributions, thus failure to reject hypothesis of equal means is more likely. In contrast, we test three different variables which increases the chance of finding a case where the test rejects hypothesis of equal means. As a countermeasure for rejecting the true hypothesis we use Bonferroni correction. 
In the end, the tests did identify two cases of different means from one case of identical means.
We note that the population who answered incorrectly to Q3 is small and recommend validation using a larger population.

\subsection{Conclusion}

In this study, we have confirmed the results of earlier studies~\cite{lehtinen2021students, santos2022jask, lehtinen2023automated}. We found similar levels of incorrect answers to simple QLCs on the second programming course as had previously been found for the first weeks of starting programming. Success in two of our QLCs predicted higher median of course points which supports previous findings and adds to the testimony that QLCs can detect factors that are relevant for learning.
Thus, we argue QLCs are a promising new question type to give early warnings on weak prerequisite knowledge, which can be taken into account in the course design. Some students might benefit from a more comprehensive recap round in the beginning of the second programming course. QLCs could complement concept inventories to provide a wider set of problems to reveal students' misconceptions that hinder learning. 

More generally, we conclude that QLCs are a promising extension to automated assessment systems. QLCs can be utilized to extend the assessment of student-created programs beyond unit testing and style linting to program comprehension~\cite{lehtinen2021lets}. If implemented well, the shift in assessment may also affect student's attitudes on the importance of understanding programs~\cite{sambell1998construction}.

The reasons why students struggle to interpret their own code are not known. Novice coders tend to make extensive use of programming forums~\cite{wongaitken2022lucky} such as Stack Overflow and might use code snippets that contain code they might not fully understand. The recent introduction of free and easily available AI-based code generation tools is likely to exacerbate the problem further. 
As students start using such tools while learning programming, the teaching should focus more on tools and effective strategies for program comprehension.





In the future, more studies are needed to analyse different types of QLCs and why some students struggle to answer them correctly. Moreover, with AI-assisted programming on the rise and student enrollments high, it is ever more timely to find automated triggers for self-explanations of code and self-regulation. QLCs hold the potential to achieve that, but further research is needed to tap into that potential.



\bibliographystyle{ACM-Reference-Format}
\balance
\bibliography{base}



\end{document}